\RequirePackage{fix-cm}
\documentclass[smallextended]{svjour3}       
\smartqed  
\usepackage{graphicx}
\usepackage{amsmath}

\begin{document}

\title{Alternative writings of classical elastodynamics equations as a first order symmetric system}

\titlerunning{Classical elastodynamics equations as a first order symmetric system}        

\author{Dimitris Sfyris}


\institute{D. Sfyris \at
              Foundation for Research and Technology (FORTH) \\
Institute of Applied and Computational Mathematics (IACM) \\
100 N. Plastira str, GR-70 013, Vassilika Vouton, Heraklion, Crete, Greece\\
              \email{dsfyris@iacm.forth.gr}}

\date{Received: date / Accepted: date}

\maketitle

\begin{abstract}

We explore alternative writings of the equations of classical elastodynamics as a first order symmetric system. In the one dimensional case we present symmetric writings with respect to: i) the velocity ($u_t$) and the displacement gradient ($u_x$), ii) the velocity and stress ($\sigma$), iii) all three quantities: the velocity, the displacement gradient and the stress, and finally iv) the momentum ($\rho_0 u_t$), the velocity, the displacement gradient and the stress. In the two dimensional case we present similar writings with respect to: i) the velocity (${\bf u}_t$) and the strain tensor ($\bf e$), ii) the velocity and the stress tensor ($\boldsymbol \sigma$), iii) all three variables $({\bf u}_t, {\bf e}, \boldsymbol \sigma)$, and finally iv) one more writing utilizing the momentum as well, i.e. $(\rho_0 {\bf u}_t, {\bf u}_t, {\bf e}, \boldsymbol \sigma)$. We accomplish our goal by judiciously using the compatibility equations as well as the momentum equation and the time differentiated constitutive law. This is done in an inverse way: we start by writing our initial equations as a first order system of the form ($\bf q$ being the vector representing the variables in each writing)
$$
	A \frac{\partial {\bf q}}{\partial t}+\sum_{i=1}^n B_i \frac{\partial {\bf q}}{\partial x_i}=0,
$$
with $n=1, 2$ depending on whether we are in 1 or 2 dimensions. We then check what are the symmetric forms of matrices $B_i$ and which combinations of the compatibility equations, the momentum equations and the time differentiated constitutive law  should be used in order the symmetric form of matrices $B_i$ to appear into the system. This "symmetrization" process alters matrix $A$ and if the resulting matrix $A$ is symmetric our goal is accomplished. Our analysis is confined to classical elastodynamics, namely geometrically and materially linear anisotropic elasticity.     

\keywords{Linear theory \and First order symmetric system \and Elastodynamics }
\subclass{74B05 \and 74B99 \and 35L02 \and 35Q74}
\end{abstract}

\section{Introduction}

When writing the system of classical elastodynamics as a linear symmetric system (\cite{Friedrichs}) it is a common practice to use the stress tensor and the velocity field as the basic variables. Such an approach requires the time differentiated constitutive law to be used as a field equation (\cite{Wilcox,Hughes-Marsden,Yakhno-Akmaz}). In \cite{Sfyris2024} we present an alternative path for writing the system of classical elastodynamics as a linear symmetric system using the displacement gradient, ${\bf u}_{\bf x}$, and the velocity, ${\bf u}_t$, as the basic variables. The motivation stem from standard procedures in linear wave equations: in order to accomplish our goal we judiciously use linear combinations of the compatibility equations.

We here explore alternative writing of the equations of classical elastodynamics as a first order symmetric system; alternative is meant in the sense that in the  one dimensional case we present symmetric writings with respect to: i) the velocity ($u_t$) and the displacement gradient ($u_x$), ii) the velocity and stress ($\sigma$), iii) all three quantities: the velocity, the displacement gradient and the stress, and finally iv)the momentum ($\rho_0 u_t$), the velocity, the displacement gradient and the stress. In the two dimensional case we present similar writings with respect to: i) the velocity (${\bf u}_t$) and the strain tensor ($\bf e$), ii) the velocity and the stress tensor ($\boldsymbol \sigma$), iii) all three variables $({\bf u}_t, {\bf e}, \boldsymbol \sigma)$, and finally iv) one more writing utilizing the momentum as well, i.e. $(\rho_0 {\bf u}_t, {\bf u}_t, {\bf e}, \boldsymbol \sigma)$. 

The strategy used to accomplish this goal is similar to that of \cite{Sfyris2024}; we work in an inverse way: we start by writing our initial equations as a first order system of the form ($\bf q$ being the vector representing the variables in each case)
\begin{equation}
	A \frac{\partial {\bf q}}{\partial t}+\sum_{i=1}^n B_i \frac{\partial {\bf q}}{\partial x_i}=0,
\end{equation}
with $n=1, 2$ depending on whether we are in 1 or 2 dimensions. We then check what are the symmetric forms of matrices $B_i$ and which combinations of the compatibility equations, the momentum equation as well as the time differentiated constitutive laws  should be used in order the symmetric form of matrices $B_i$ to appear into the system. This "symmetrization" process alters matrix $A$ and if the resulting matrix $A$ is symmetric our goal is accomplished.

This way we find symmetric writings for the one dimensional case using as unknowns the pair of velocity and the displacement gradient; also the pair of velocity and stress. These works are common more or less in the classical elasticity literature. The first one writing uses the compatibility equation multiplied judiciously (\cite{Sfyris2024}), while the second writing utilizes the constitutive law time differentiated (\cite{Wilcox,Hughes-Marsden,Yakhno-Akmaz}). We combine these two writings and find a symmetric writing with respect to all three variables: velocity, displacement gradient and stress. To accomplish such a goal we use the momentum equation, the compatibility relation as well as the time differentiated constitutive law. It is important to stress that we add the momentum equation once written with respect to the stress and once written with respect to the displacement gradient in order to achieve our purpose. Such an approach that utilizes all three variables might help when characterizing, in terms of the hyperbolicity, implicit theories (\cite{Sfyrisetal2024}). 

Even though quite formal, we present writings using the momentum instead of the velocity motivated from the fact that these two quantities (velocity and momentum) are conjugate in the sense that they produce the kinetic energy of the system. We offer writings using the  momentum and the stress, the momentum and the displacement gradient as well as a writing using the momentum, the stress and the displacement gradient. For this last writing we have to use twice the momentum; once written with respect to the stress and once written with respect to the displacement gradient and add them together.

We also present a very general writing using the momentum, the velocity, the displacement gradient as well as the stress as variables. To achieve this goal we use two times the momentum equation; each of these two equations is the sum of the momentum equation once written with respect to the momentum and once with respect to the velocity. Something similar happens for the rest equations used. For the compatibility equation we write it once with respect to the momentum and once with respect to the velocity and add them together. A similar plan is used for the time differentiated constitutive law; we write it once with respect to the momentum and once with respect to the velocity and add the outcome. This way we built symmetric writings of our system. 

In the two dimensional case an analogous program is put forth. We present writings using the strain tensor and the velocity, the stress tensor and the velocity as well a writing using the stress tensor, the strain tensor and the velocity. We can also, quite formally, substitute the velocity by the momentum. We also present a writing where all four quantities: velocity, momentum, stress and strain tensor are present. To infer which equations should be used we are guided by the one dimensional case. We avoid doing the three dimensional case, since the calculations are straighforard generalizations of the two dimensional case. 

The methodology used here is motivated from the strategy used in \cite{Sfyris2024}. There, we present, essentially, an inverse way of building a symmetric system. We write the system down and examine which terms should be added in order the matrices related with the space derivatives are symmetric. We then check what happens with the matrix related with the time derivative. This inverse way of working results ultimately in a symmetric system, since all matrices present are symmetric. 

The article is structured as follows. In Section 2 we focus on the one dimensional case and offer symmetric writings with respect to variables $(u_t, u_x)$, $(u_t, \sigma)$, ($u_t, u_x, \sigma$), $(\rho_0 u_t, u_x)$, $(\rho_0 u_t, \sigma)$, ($\rho_0 u_t, u_x, \sigma$) and ($\rho_0 u_t, u_t, u_x, \sigma$). In Section 3 we treat the two dimensional case and present symmetric writings with respect to $({\bf u}_t, {\bf e})$, (${\bf u}_t, \boldsymbol \sigma$), (${\bf u}_t, {\bf e}, {\boldsymbol \sigma}$) and ($\rho_0 {\bf u}_t, {\bf u}_t, {\bf e}, {\boldsymbol \sigma}$). The article ends up with some concluding remarks and potential future directions in Section 4. For the one dimensional case we denote space differentiation as $()_x=\frac{\partial ()}{\partial x}$ and time differentiation as $()_t=\dot{()}=\frac{\partial ()}{\partial t}$ for a quantity $()$. In the two dimensional case we use instead $()_{,i}=\frac{\partial ()}{\partial x_i}$ and $()_{,t}=\dot{()}=\frac{\partial ()}{\partial t}$.     

\section{Elasticity in 1d}

In the one dimensional setting the strain component is essentially the displacement gradient ($u_x$). The starting point is the elastic energy which has the form 
\begin{equation}
W(u_x)=\frac{1}{2} \alpha u_x^2,
\end{equation}
$\alpha$ being the material parameter of the model. The momentum equation in the absence of body forces and with a unit density reads
\begin{equation}
u_{tt}-\sigma_x=0,
\end{equation}
while the constitutive law is
\begin{equation}
\sigma=\frac{\partial W}{\partial u_x}.
\end{equation}
When the material density is not unity, one has to add the term $\rho_0$ in the acceleration term on the left hand side of eq. (3). 

In the next subsection we present writings using $(u_t, u_x)$ following \cite{Sfyris2024}, $(u_t, \sigma)$ following \cite{Yakhno-Akmaz}, and then generalize to writings using ($u_t, u_x, \sigma$), $(\rho_0 u_t, u_x)$, $(\rho_0 u_t, \sigma)$, ($\rho_0 u_t, u_x, \sigma$) and ($\rho_0 u_t, u_t, u_x, \sigma$).

\subsection{Elasticity in 1d with respect to $(u_t, u_x)$}

The momentum equation gives
\begin{equation}
u_{tt}=\sigma_x= \alpha u_{xx} + \alpha_x u_x.
\end{equation} 
Setting $u_t=v, u_x=e$ the momentum equation renders
\begin{equation}
v_t=\alpha e_x +\alpha_x e, 
\end{equation} 
while the compatibility equation renders
\begin{equation}
v_x=e_t \rightarrow v_x - e_t =0.
\end{equation} 
We multiply the compatibility relation by $\alpha$ to obtain   
\begin{equation}
\alpha v_x - \alpha  e_t =0.
\end{equation} 
So, as a system eqs. (6, 8) are 
\begin{eqnarray}
v_t-\alpha e_x &&= \alpha_x e, \\
\alpha e_t-\alpha v_x &&=0.
\end{eqnarray}
This system can be written in the form 
\begin{equation}
A_{\text{sym}}^{(u_t, u_x)} \frac{\partial {\bf q}}{\partial t} + B_{\text{sym}}^{(u_t, u_x)} \frac{\partial {\bf q}}{\partial x}=f^{(u_t, u_x)},
\end{equation} 
with respect to ${\bf q}=(v, e)^T$ with 
\begin{equation}
A_{\text{sym}}^{(u_t, u_x)}=\begin{pmatrix}
1 & 0  \\
0 & \alpha  
\end{pmatrix},
B_{\text{sym}}^{(u_t, u_x)}=\begin{pmatrix}
0 & -\alpha  \\
-\alpha & 0  
\end{pmatrix},
f^{(u_t, u_x)}=\begin{pmatrix}
\alpha_x e  \\
0  
\end{pmatrix}.
\end{equation}

Matrices $A_{\text{sym}}^{(u_t, u_x)}$ and $B_{\text{sym}}^{(u_t, u_x)}$ are symmetric (due to the suitable multiplication of the compatibility equation wth $\alpha$), so our system is put into symmetric format. The eigenvalues of the matrix $A_{\text{sym}}^{(u_t, u_x)}$ are $(1, \alpha)$ so, when $\alpha > 0$ the matrix $A_{\text{sym}}^{(u_t, u_x)}$ is positive definite.  

\subsection{Elasticity in 1d with respect to $(u_t, \sigma)$}

Using again as a starting point the energy of eq. (2), stress is given by eq. (4). Now, we need to time differentiate the constitutive law. If this is done directly in eq. (4) one obtains
\begin{equation}
\dot{\sigma}=\dot{\alpha} u_x+\alpha \dot{u}_x. 
\end{equation} 
If we set $u_t=v$ then the first term in the right hand side of the last equation cannot be transformed to the new system of variables, $(u_t, \sigma)$. To by-pass this issue we write the constitutive law in the form 
\begin{equation}
\alpha^{-1} \sigma= u_x,  
\end{equation} 
which requires that the material parameter $\alpha$ can be inverted (i.e $\alpha \neq 0$). By time differentiating the last equation we obtain 
\begin{equation}
\alpha^{-1}_t \sigma+\alpha^{-1} \sigma_t= \dot{u}_x.   
\end{equation} 
When we set $u_t=v$ this equation can be written in the new system as
\begin{equation}
\alpha^{-1}_t \sigma+\alpha^{-1} \sigma_t= v_x.   
\end{equation} 
If to this last equation we add the momentum equation we are essentially working with the system 
\begin{eqnarray}
v_t-\sigma_x &&= 0, \\
\alpha^{-1} \sigma_t- v_x &&=-\dot{\alpha}^{-1} \sigma.
\end{eqnarray}
This system can be written in the form 
\begin{equation}
A_{\text{sym}}^{(u_t, \sigma)} \frac{\partial {\bf q}}{\partial t} + B_{\text{sym}}^{(u_t, \sigma)} \frac{\partial {\bf q}}{\partial x}=f^{(u_t, \sigma)},
\end{equation} 
with respect to ${\bf q}=(v, \sigma)^T$ with 
\begin{equation}
A_{\text{sym}}^{(u_t, \sigma)}=\begin{pmatrix}
1 & 0  \\
0 & \alpha^{-1}  
\end{pmatrix},
B_{\text{sym}}^{(u_t, \sigma)}=\begin{pmatrix}
0 & -1  \\
-1 & 0  
\end{pmatrix},
f^{(u_t, \sigma)}=\begin{pmatrix}
0  \\
-\dot{\alpha}^{-1} \sigma  
\end{pmatrix}.
\end{equation}

Matrices $A_{\text{sym}}^{(u_t, \sigma)}$ and $B_{\text{sym}}^{(u_t, \sigma)}$ are symmetric, so our system is put into symmetric format. Eigenvalues of the matrix $A_{\text{sym}}^{(u_t, \sigma)}$ are $(1, \frac{1}{\alpha})$ so, when $\alpha > 0$ matrix $A_{\text{sym}}^{(u_t,\sigma)}$ is positive definite.

\subsection{Elasticity in 1d with respect to $(u_t, u_x, \sigma)$}

On the one hand, the writing with respect to $(u_t, u_x)$ utilizes the compatibility equation and the momentum equation. On the other hand, the writing with respect to $(u_t, \sigma)$ uses the momentum equation and the constitutive law time differentiated. So, a natural question arises on whether there is a symmetric writing using all three variables $(u_t, u_x, \sigma)$. Such a writing  might help in the classification, in terms of the hyperbolicity, in implicit theories where the constitutive law is given in implicit form (see \cite{Sfyrisetal2024}). 

If the momentum equation is expressed in terms of $u_x$, the compatibility equation is multiplied by $\alpha$ and we also use the time differentiated constitutive law, we end up working with equations  
\begin{eqnarray}
v_t-\alpha e_x &&= \alpha_x e, \\
\alpha e_t-\alpha v_x &&=0, \\
\alpha^{-1} \sigma_t- v_x &&=-\dot{\alpha}^{-1} \sigma.
\end{eqnarray}
This system can be written in the form 
\begin{equation}
A^{(u_t, e, \sigma)} \frac{\partial {\bf q}}{\partial t} + B^{(u_t, e, \sigma)} \frac{\partial {\bf q}}{\partial x}=f^{(u_t, e, \sigma)},
\end{equation} 
with respect to ${\bf q}=(v, e, \sigma)^T$ with 
\begin{equation}
A^{(u_t, e, \sigma)}=\begin{pmatrix}
1 & 0  & 0 \\
0 & \alpha & 0 \\
0 & 0 & \alpha^{-1}  
\end{pmatrix},
B^{(u_t, e, \sigma)}=\begin{pmatrix}
0 & -\alpha & 0  \\
-\alpha & 0 & 0 \\
-1 & 0 & 0  
\end{pmatrix},
\end{equation}
$B^{(u_t, e, \sigma)}$ being obviously not symmetric. In order to make the last matrix symmetric one has to add to the (1, 3)-slot of the matrix $B^{(u_t, e, \sigma)}$ the term -1 which corresponds to term -$\sigma_x$. Such term can be found in the momentum equation when it is written with respect to the stress, namely 
\begin{equation}
v_t-\sigma_x=0.
\end{equation}

So, if we add the two expressions of the momentum equations (once written with respect to $\sigma$ and once with respect to $u_x$) 
\begin{eqnarray}
v_t-\sigma_x&&=0. \\
v_t-\alpha e_x &&= \alpha_x e,
\end{eqnarray}
we work with 
\begin{equation}
2v_t-\alpha e_x-\sigma_x =\alpha_x e.
\end{equation}
So, in this case we end up working with equations
\begin{eqnarray}
2v_t-\alpha e_x-\sigma_x &&=\alpha_x e, \\
\alpha e_t-\alpha v_x &&=0, \\
\alpha^{-1} \sigma_t- v_x &&=-\dot{\alpha}^{-1} \sigma.
\end{eqnarray}
These equations can be written as a system in the form 
\begin{equation}
A_{\text{sym}}^{(u_t, e, \sigma)} \frac{\partial {\bf q}}{\partial t} + B_{\text{sym}}^{(u_t, e, \sigma)} \frac{\partial {\bf q}}{\partial x}=f^{(u_t, e, \sigma)},
\end{equation} 
with respect to ${\bf q}=(v, e, \sigma)^T$ with 
\begin{equation}
A_{\text{sym}}^{(u_t, e, \sigma)}=\begin{pmatrix}
2 & 0  & 0 \\
0 & \alpha & 0 \\
0 & 0 & \alpha^{-1}  
\end{pmatrix},
B_{\text{sym}}^{(u_t, e, \sigma)}=\begin{pmatrix}
0 & -\alpha & -1  \\
-\alpha & 0 & 0 \\
-1 & 0 & 0  
\end{pmatrix},
f^{(u_t, e, \sigma)}=\begin{pmatrix}
\alpha_x e \\
0  \\
-\dot{\alpha}^{-1} \sigma  
\end{pmatrix}
\end{equation}

Matrices $A_{\text{sym}}^{(u_t, e, \sigma)}$ and $B_{\text{sym}}^{(u_t, e, \sigma)}$ are symmetric so our symmetric writing is achieved. We stress that to accomplish this goal we used the momentum equation, the compatibility equation (multiplied judiciously) as well as the time differentiated constitutive law. For the momentum equation used, we add the momentum equation written once with respect to $u_x$ and once with respect to $\sigma$. Matrix $A_{\text{sym}}^{(u_t, e, \sigma)}$ has determinant 2 and eigenvalues $(1, \alpha, \frac{1}{\alpha})$, so when $\alpha >0$ the matrix $A_{\text{sym}}^{(u_t, e, \sigma)}$ is positive definite.

\subsection{Elasticity in 1d with respect to $(\rho_0 u_t, u_x)$}

Even though it is formal, we seek for symmetric writings with respect to the momentum $\rho_0 u_t$. Such an analysis stem from the fact that the momentum and the velocity are conjugate in the sense that they produce the kinetic energy ($\frac{1}{2} \rho_0 {\bf u}_t {\bf u}_t$) in the same spirit that stress and strain are conjugate. This essentially means that the mass density is non-trivial, nevertheless it remains constant in space and time. In this case the momentum equation reads
\begin{equation}
\rho_0 u_{tt} = \alpha u_{xx} + \alpha_x u_x. 
\end{equation}
Setting $\rho_o u_t=v, u_x =e$ we have for the momentum equation
\begin{equation}
v_t = \alpha e_x + \alpha_x e. 
\end{equation}
The compatibility equation multiplied by $\rho_0$ reads
\begin{equation}
\rho_0 e_t = v_x. 
\end{equation}
If we choose to work with the last two equations our system will not be in  symmetric form. In order to bring it into symmetric form we multiply the compatibility equation with $\alpha$ and finally work with 
\begin{eqnarray}
v_t-\alpha e_x &&= \alpha_x e \\
\rho_0 \alpha e_t-\alpha e_x && =0,
\end{eqnarray}
which renders a symmetric writing in the form 
\begin{equation}
A_{\text{sym}}^{(\rho_0 u_t, u_x)} \frac{\partial {\bf q}}{\partial t} + B_{\text{sym}}^{(u_t, u_x)} \frac{\partial {\bf q}}{\partial x}=f^{(u_t, u_x)},
\end{equation} 
with respect to ${\bf q}=(\rho_0 u_t, e)^T$ with 
\begin{equation}
A_{\text{sym}}^{(\rho_0 u_t, u_x)}=\begin{pmatrix}
1 & 0  \\
0 & \rho_0 \alpha  
\end{pmatrix},
B_{\text{sym}}^{(\rho_0 u_t, u_x)}=\begin{pmatrix}
0 & -\alpha  \\
-\alpha & 0  
\end{pmatrix},
f^{(\rho_0 u_t, u_x)}=\begin{pmatrix}
\alpha_x e  \\
0  
\end{pmatrix}.
\end{equation}
Obviously, matrices $A^{(\rho_0 u_t, u_x)}_{\text{sym}}$, $B^{(\rho_0 u_t, u_x)}_{\text{sym}}$ are symmetric and when $\rho_0, \alpha >0$, $A^{(\rho_0 u_t, u_x)}_{\text{sym}}$ is positive definite. 

\subsection{Elasticity in 1d with respect to $(\rho_0 u_t, \sigma)$}

The time differentiated constitutive law multiplied by $\rho_0$ has the form
\begin{equation}
\rho_0\alpha^{-1}_t \sigma+\rho_0\alpha^{-1} \sigma_t= \rho_0v_x.   
\end{equation} 
If to this last equation we add the momentum equation we are essentially working with the system 
\begin{eqnarray}
v_t-\sigma_x &&= 0, \\
\rho_0\alpha^{-1} \sigma_t- v_x &&=-\rho_0\dot{\alpha}^{-1} \sigma.
\end{eqnarray}
This system can be written in the form 
\begin{equation}
A_{\text{sym}}^{(\rho_0 u_t, \sigma)} \frac{\partial {\bf q}}{\partial t} + B_{\text{sym}}^{(\rho_0u_t, \sigma)} \frac{\partial {\bf q}}{\partial x}=f^{(\rho_0u_t, \sigma)},
\end{equation} 
with respect to ${\bf q}=(v, \sigma)^T$ with 
\begin{equation}
A_{\text{sym}}^{(\rho_0u_t, \sigma)}=\begin{pmatrix}
1 & 0  \\
0 & \alpha^{-1} \rho_0  
\end{pmatrix},
B_{\text{sym}}^{(\rho_0 u_t, \sigma)}=\begin{pmatrix}
0 & -1  \\
-1 & 0  
\end{pmatrix},
f^{(u_t, \sigma)}=\begin{pmatrix}
0  \\
-\dot{\alpha}^{-1} \rho_0 \sigma  
\end{pmatrix},
\end{equation}
which is in symmetric format since matrices $A_{\text{sym}}^{(\rho_0u_t, \sigma)}$, $B_{\text{sym}}^{(\rho_0u_t, \sigma)}$ are symmetric.

\subsection{Elasticity in 1d with respect to $(\rho_0 u_t, u_x, \sigma)$}

In analogy with the writing with respect to $(u_t, u_x, \sigma)$, we work with the momentum equation, the compatibility equation and the constitutive law time differentiated. Setting $\rho_0 u_t =v, u_x=e$ and letting $\sigma$ as is, we work with the momentum equation, the constitutive law multiplied by $\rho_0 \alpha$ and the time differentiated constitutive law multiplied by $\alpha$
\begin{eqnarray}
v_t-\alpha e_x &&= \alpha_x e, \\
\rho_0 \alpha e_t-\alpha v_x &&=0, \\
\rho_0 \alpha^{-1} \sigma_t- v_x &&=-\dot{\alpha}^{-1} \rho_0 \sigma.
\end{eqnarray}
This system can be written in the form 
\begin{equation}
A^{(\rho_0 u_t, e, \sigma)} \frac{\partial {\bf q}}{\partial t} + B^{(\rho_0 u_t, e, \sigma)} \frac{\partial {\bf q}}{\partial x}=f^{(\rho_0 u_t, e, \sigma)},
\end{equation} 
with respect to ${\bf q}=(v, e, \sigma)^T$ with 
\begin{equation}
A^{(\rho_0 u_t, u_x, \sigma)}=\begin{pmatrix}
1 & 0  & 0 \\
0 & \rho_0 \alpha & 0 \\
0 & 0 & \rho_0 \alpha^{-1}  
\end{pmatrix},
B^{(\rho_0u_t, u_x, \sigma)}=\begin{pmatrix}
0 & -\alpha & 0  \\
-\alpha & 0 & 0 \\
-1 & 0 & 0  
\end{pmatrix},
\end{equation}
\begin{equation}
f^{(\rho_0 u_t, \sigma)}=\begin{pmatrix}
\alpha_x e  \\
0  \\
-\dot{\alpha}^{-1} \rho_0 \sigma
\end{pmatrix}.
\end{equation}
which is obviously not in symmetric format since the matrix $B^{(\rho_0u_t, u_x, \sigma)}$ is not symmetric. To make this matrix symmetric we should add in its (1, 3)-slot the term -1, which corresponds to term $-\sigma_x$. This term can be brought about from the momentum equation when written with respect to stresses: $v_t-\sigma_x=0$. Adding this equation to the momentum equation written with respect to $u_x$ we finally work with 
\begin{eqnarray}
2v_t-\alpha e_x -\sigma_x &&= \alpha_x e, \\
\rho_0 \alpha e_t-\alpha v_x &&=0, \\
\rho_0 \alpha^{-1} \sigma_t- v_x &&=-\dot{\alpha}^{-1} \rho_0 \sigma.
\end{eqnarray}
This system can be written in the form 
\begin{equation}
A_{\text{sym}}^{(\rho_0 u_t, e, \sigma)} \frac{\partial {\bf q}}{\partial t} + B_{\text{sym}}^{(\rho_0 u_t, e, \sigma)} \frac{\partial {\bf q}}{\partial x}=f^{(\rho_0 u_t, e, \sigma)},
\end{equation} 
with respect to ${\bf q}=(v, e, \sigma)^T$ with 
\begin{equation}
A_{\text{sym}}^{(\rho_0u_t, e, \sigma)}=\begin{pmatrix}
2 & 0  & 0 \\
0 & \rho_0 \alpha & 0 \\
0 & 0 & \rho_0 \alpha^{-1}  
\end{pmatrix},
B_{\text{sym}}^{(\rho_0u_t, \sigma)}=\begin{pmatrix}
0 & -\alpha & -1  \\
-\alpha & 0 & 0 \\
-1 & 0 & 0  
\end{pmatrix},
\end{equation}
\begin{equation}
f^{(\rho_0 u_t, \sigma)}=\begin{pmatrix}
\alpha_x e  \\
0  \\
-\dot{\alpha}^{-1} \rho_0 \sigma
\end{pmatrix}.
\end{equation}
In this format the system is put into symmetric form. Matrix $A^{(\rho_0u_t, e, \sigma)}$ has determinant $2\rho_0^2>0$, so it can be inverted and eigenvalues $(2, \frac{\rho_0}{\alpha}, \rho_0\alpha)$ so, when $\alpha>0$ and $\rho_0>0$ it is positive definite.  

\subsection{Elasticity in 1d with respect to $(\rho_0 u_t, u_t, u_x, \sigma)$}

In order to write the system of equations in this format we set $v^1=\rho_0 u_t, v^2=u_t$. We use the momentum equations twice: once written with respect to $v^1$ and once with respect to $v^2$. In each one of these equations we should add the momentum equation again expressed with respect to stresses. We also write the compatibility once with respect to $v^1$ and once with respect to $v^2$ and add them. The same is done for the time differentiated constitutive law: we write it once with respect to $v^1$ and once with respect to $v^2$ and add the outcome. This way, at the end of the day, we work with equations
\begin{eqnarray}
2v^1_t-\alpha e_x -\sigma_x &&= \alpha_x e, \\
2 \rho_0 v^2_t-\alpha e_x -\sigma_x &&= \alpha_x e, \\
\alpha(\rho_0+1) e_t-\alpha v^1_x-\alpha v^2_x &&=0, \\
\alpha^{-1}(\rho_0+1) \sigma_t- v^1_x-v^2_x &&=-\dot{\alpha}^{-1} (\rho_0+1) \sigma.
\end{eqnarray}
Eq. (59) is twice the momentum equation: with respect to $v^1$ for the inertial term and once with respect to $u_x$ and once with respect to $\sigma$ for the other term. Similar approach holds for eq. (60): the momentum equation is used twice using $v^2$ for the inertial term and for the other term once we use $u_x$ and once $\sigma$. Eq. (61) is the addition of the compatibility equation once written with respect to $v^1$ and once with respect to $v^2$ multiplied judiciously. Eq. (62) stem from the addition of the time differentiated constitutive law once written with respect to $v^1$ and once with respect to $v^2$.

This system can be written in the form 
\begin{equation}
A_{\text{sym}}^{(\rho_0 u_t, u,_t, e, \sigma)} \frac{\partial {\bf q}}{\partial t} + B_{\text{sym}}^{(\rho_0 u_t, u_t, e, \sigma)} \frac{\partial {\bf q}}{\partial x}=f^{(\rho_0 u_t, u_t, e, \sigma)},
\end{equation} 
with respect to ${\bf q}=(v^1, v^2, e, \sigma)^T$ with 
\begin{equation}
A_{\text{sym}}^{(\rho_0 u_t, u_t, e, \sigma)}=\begin{pmatrix}
2 & 0  & 0 & 0\\
0 & 2 \rho_0  & 0 & 0 \\
0 & 0 & \alpha (\rho_0+1) & 0 \\
0 & 0 & 0 & \alpha^{-1} (\rho_0+1)   
\end{pmatrix},
\end{equation}
\begin{equation}
B_{\text{sym}}^{(\rho_0 u_t, u_t, e, \sigma)}=\begin{pmatrix}
0 & 0 &  -\alpha & -1  \\
0 & 0 &  -\alpha & -1  \\
-\alpha & -\alpha & 0 & 0 \\
-1 & -1 & 0 & 0  
\end{pmatrix},
\end{equation}
\begin{equation}
f_{\text{sym}}^{(\rho_0 u_t, u_t, e, \sigma)}=\begin{pmatrix}
\alpha_x e  \\
\alpha_x e  \\
0  \\
-\dot{\alpha}^{-1} \rho_0 \sigma
\end{pmatrix}.
\end{equation}
In this format the system is put into symmetric form. Matrix $A_{\text{sym}}^{(\rho_0 u_t, u_t, e, \sigma)}$ has determinant $4 \rho_0 (\rho_0+1)^2>0$, so it is invertible and eigenvalues $(2, 2\rho_0, \frac{1+\rho_0}{\alpha}, \alpha(\rho_0+1))$ so it positive definite when $\rho_0, \alpha>0$.

\section{Elasticity in 2d}

In the two dimensional case the strain tensor $\bf e$ is related with the displacement gradient according to 
\begin{equation}
	e_{ij}=\frac{1}{2}\left( u_{i, j} + u_{j,i}   \right), i, j=1, 2.
\end{equation}
The starting point is the momentum equation which, in the absence of body forces and with a unit density, has the form 
\begin{equation}
	\ddot{u}_i = \sigma_{ij,j},
\end{equation}
where $\sigma$ is the classical Cauchy stress tensor. If the mass density is not unity, then a term $\rho_0$ should be present in the left hand side of the last equation. Classical elasticity utilizes the energy 
\begin{equation}
	W=\frac{1}{2} e_{ij} \mathcal{C}_{ijkl} e_{kl},
\end{equation}
where $\mathcal{C}$ are the elasticities of a generically anisotropic material which satisfy minor and major symmetries $\mathcal{C}_{ijkl}=\mathcal{C}_{jikl}=\mathcal{C}_{ijlk}=\mathcal{C}_{klij}$. The stress tensor is then 
\begin{equation}
	\sigma_{ij}=\frac{\partial W}{\partial e_{ij}}=\mathcal{C}_{ijkl} e_{kl},
\end{equation} 
so the momentum equation reads
\begin{equation}
	\ddot{u}_i = \mathcal{C}_{ijkl} e_{kl,j}.
\end{equation}

If one furthermore uses eq. (67), then the momentum equation written with respect to the displacements, has the form 
\begin{eqnarray}
	&& u_{i,tt} = \frac{1}{2} \mathcal{C}_{ijkl} ( u_{k,lj} + u_{l,kj} ) \rightarrow \nonumber\\
	&& u_{i,tt} - \frac{1}{2} \mathcal{C}_{ijkl}  u_{k,lj} - \frac{1}{2} \mathcal{C}_{ijkl}  u_{l,kj} = 0. 
\end{eqnarray}
In the next subsection we use eqs. (72) together with the compatibility equation and the time differentiated law in order to present alternative writings of the classical elastodynamics equations as a first order linear symmetric system in the form 
\begin{equation}
A_{\text{sym}} \frac{\partial {\bf q}}{\partial t}+\sum_{i=1}^2 B_{i \  {\text{sym}}} \frac{\partial {\bf q}}{\partial x_i}=0.
\end{equation}
Matrix $B_{1 \  {\text{sym}}}$ collects terms related with derivatives with respect to $x_1$ while matrix $B_{2 \  {\text{sym}}}$ collects terms related with derivatives with respect to $x_2$. 

\subsection{Elasticity in 2D with respect to $({\bf u}_t, {\bf e})$}

In the two dimensional case the first of the momentum equation reads
\begin{eqnarray}
u_{1,tt}&&-{\mathcal C}_{1111} u_{1,11}-{\mathcal C}_{1211} u_{1,12}-{\mathcal C}_{1112} u_{1,21}-{\mathcal C}_{1212} u_{1,22} \nonumber\\
&& -{\mathcal C}_{1112} u_{2,11}-{\mathcal C}_{1212} u_{2,12}-{\mathcal C}_{1122} u_{2,21}-{\mathcal C}_{1222} u_{2,22}=0,
\end{eqnarray}
which we write in the following form in order to built the first strain component
\begin{eqnarray}
u_{1,tt}&&-{\mathcal C}_{1111} u_{1,11}-{\mathcal C}_{1211} u_{1,12}-{\mathcal C}_{1112} (u_{1,2}+u_{2,1})_{,1} \nonumber\\
&&-{\mathcal C}_{1212}( u_{1,2}+u_{2,1})_{,2}-{\mathcal C}_{1122} u_{2,21}-{\mathcal C}_{1222} u_{2,22}=0,
\end{eqnarray}
Now if we set
\begin{eqnarray}
{\bf q}=&&\{v^1=e_{11}=u_{1,1}, v^2=\frac{1}{2}(u_{1,2}+u_{2,1})=e_{12}, \nonumber\\ 
&& v^3=u_{1,t}, v^4=e_{22}=u_{2,2}, v^5=u_{2,t}  \}
\end{eqnarray}
the momentum equation reads
\begin{eqnarray}
v^3_{1,t}&&-{\mathcal C}_{1111} v^1_{,1}-{\mathcal C}_{1211} v^1_{,2}-{\mathcal C}_{1112} 2 v^2_{,1} \nonumber\\
&&-{\mathcal C}_{1212} 2v^2_{,2}-{\mathcal C}_{1122} v^4_{,1}-{\mathcal C}_{1222} v^4_{,2}=0,
\end{eqnarray}
and this is the first of the momentum equations written with respect to the new system. 

The second of the momentum equation reads
\begin{eqnarray}
u_{2,tt}&&-{\mathcal C}_{2111} u_{1,11}-{\mathcal C}_{2211} u_{1,12}-{\mathcal C}_{2112} u_{1,21}-{\mathcal C}_{2212} u_{1,22} \nonumber\\
&& -{\mathcal C}_{2112} u_{2,11}-{\mathcal C}_{2212} u_{2,12}-{\mathcal C}_{2122} u_{2,21}-{\mathcal C}_{2222} u_{2,22}=0,
\end{eqnarray}
which can also be written in the form
\begin{eqnarray}
u_{2,tt}&&-{\mathcal C}_{2111} u_{1,11}-{\mathcal C}_{2211} u_{1,12}-{\mathcal C}_{2112} (u_{1,2}+u_{2,1})_{,1} \nonumber\\
&&-{\mathcal C}_{2212}( u_{1,2}+u_{2,1})_{,2}-{\mathcal C}_{2122} u_{2,21}-{\mathcal C}_{2222} u_{2,22}=0,
\end{eqnarray}
so after the setting of eq. (76) takes the form
\begin{eqnarray}
v^5_{1,t}&&-{\mathcal C}_{2111} v^1_{,1}-{\mathcal C}_{2211} v^1_{,2}-{\mathcal C}_{2112} 2 v^2_{,1} \nonumber\\
&&-{\mathcal C}_{2212} 2v^2_{,2}-{\mathcal C}_{2122} v^4_{,1}-{\mathcal C}_{2222} v^4_{,2}=0,
\end{eqnarray}
and this is the second of the momentum equations written with respect to the new system. 

The compatibility relations are
\begin{eqnarray}
u_{1,1t}=&& u_{1,t1}, \\
u_{2,2t}=&& u_{2,t2}, 
\end{eqnarray}
as well as 
\begin{eqnarray}
u_{1,2t}=&& u_{1,t2}, \\
u_{2,1t}=&& u_{2,t1}, 
\end{eqnarray}
which should be added in order to built up $e_{12}=v^2$
\begin{equation}
(u_{1,2}+u_{2,1})_{,t}=u_{1,t2}+u_{2,t1}.
\end{equation}
So, all in all, the compatibility equations written with respect to the new variables are
\begin{eqnarray}
v^1_{,t}-v^3_{,1}=&&0, \\
v^4_{,t}-v^5_{,2}=&&0, \\
2v^2_{,t}-v^3_{,2}-v^5_{,1}=&&0
\end{eqnarray}

Now, if we use the compatibility and the momentum equations in the order eq. (86, 88, 77, 87, 80) then these equations can be written in the form
\begin{equation}
A^{({\bf u}_t, {\bf e})} \frac{\partial {\bf q}}{\partial t} + \sum_{i=1}^2 B_i^{({\bf u}_t, {\bf e})} \frac{\partial {\bf q}}{\partial x_i}={\bf 0},
\end{equation} 
where
\begin{equation}
A^{({\bf u}_t, {\bf e})}=\begin{pmatrix}
1 & 0  & 0 & 0 & 0 \\
0 & 2  & 0 & 0 & 0 \\
0 & 0 & 1 & 0 & 0 \\
0 & 0 & 0 & 1 & 0 \\
0 & 0 & 0 & 0 & 1 
\end{pmatrix},
\end{equation}
and
\begin{equation}
B_1^{({\bf u}_t, {\bf e})}=\begin{pmatrix}
0 & 0  & -1 & 0 & 0 \\
0 & 0  & 0 & 0 & -1 \\
-{\mathcal C}_{1111} & -2{\mathcal C}_{1112} & 0 & -{\mathcal C}_{1122} & 0 \\
0 & 0 & 0 & 0 & 0 \\
-{\mathcal C}_{2111} & -2{\mathcal C}_{2112} & 0 & -{\mathcal C}_{2122} & 0 
\end{pmatrix},
\end{equation}
as well as 
\begin{equation}
B_2^{({\bf u}_t, {\bf e})}=\begin{pmatrix}
0 & 0  & 0 & 0 & 0 \\
0 & 0  & -1 & 0 & 0 \\
-{\mathcal C}_{1211} & -2{\mathcal C}_{2112} & 0 & -{\mathcal C}_{1222} & 0 \\
0 & 0 & 0 & 0 & -1 \\
-{\mathcal C}_{2211} & -2{\mathcal C}_{2212} & 0 & -{\mathcal C}_{2222} & 0 
\end{pmatrix},
\end{equation}
when major and minor symmetries are enforced. The above matrices are clearly not symmetric.  

To proceed, we follow a procedure similar to that of \cite{Sfyris2024} in order to make the system symmetric. The symmetric form of the matrices  $B_i^{({\bf u}_t, {\bf e})}, i=1, 2$ is 
\begin{equation}
B_{1 \ \  \text{sym}}^{({\bf u}_t, {\bf e})}=\begin{pmatrix}
0 & 0  & -{\mathcal C}_{1111} & 0 & -{\mathcal C}_{2111} \\
0 & 0  & -2{\mathcal C}_{1112} & 0 & -2{\mathcal C}_{2112} \\
-{\mathcal C}_{1111} & -2{\mathcal C}_{1112} & 0 & -{\mathcal C}_{1122} & 0 \\
0 & 0 & -{\mathcal C}_{1122} & 0 & -{\mathcal C}_{2122} \\
-{\mathcal C}_{2111} & -2{\mathcal C}_{2112} & 0 & -{\mathcal C}_{2122} & 0 
\end{pmatrix},
\end{equation}
as well as 
\begin{equation}
B_{2 \ \  \text{sym}}^{({\bf u}_t, {\bf e})}=\begin{pmatrix}
0 & 0  & -{\mathcal C}_{1211} & 0 & -{\mathcal C}_{2211} \\
0 & 0  & -2{\mathcal C}_{1212} & 0 & -2{\mathcal C}_{2212} \\
-{\mathcal C}_{1211} & -2{\mathcal C}_{2112} & 0 & -{\mathcal C}_{1222} & 0 \\
0 & 0 & -{\mathcal C}_{1222} & 0 & -{\mathcal C}_{2222} \\
-{\mathcal C}_{2211} & -2{\mathcal C}_{2212} & 0 & -{\mathcal C}_{2222} & 0 
\end{pmatrix}.
\end{equation} 
We have to check what this symmetrization process adds to the matrix $A^{({\bf u}_t, {\bf e})}$. In contrast to the analysis of \cite{Sfyris2024}, we here should seek in tandem the equations that render the system symmetric. Namely, we should work with both  $B_{i \ \  \text{sym}}^{({\bf u}_t, {\bf e})}, i=1, 2$ together. The initial compatibility equation is 
\begin{equation}
v^1_{,t}-v^3_{,1}=0, 
\end{equation}
which is multiplied with ${\mathcal C}_{1111}$ in order to build the third slot of the first line of $B_{1 \ \  \text{sym}}^{({\bf u}_t, {\bf e})}$
\begin{equation}
{\mathcal C}_{1111} v^1_{,t} - {\mathcal C}_{1111} v^3_{,1}=0.
\end{equation}
For the (1, 5)-element of $B_{1 \ \  \text{sym}}^{({\bf u}_t, {\bf e})}$ we use
\begin{equation}
2 {\mathcal C}_{2111} v^2_{,t} - {\mathcal C}_{2111} v^3_{,2}- {\mathcal C}_{2111} v^5_{,1}=0.
\end{equation}
The second term in the last equation tackles also the (1, 3)-element of $B_{2 \ \  \text{sym}}^{({\bf u}_t, {\bf e})}$ with the help of the minor symmetries. What it remains is the (1, 5)-element of $B_{1 \ \  \text{sym}}^{({\bf u}_t, {\bf e})}$. For this 
\begin{equation}
{\mathcal C}_{2211} v^4_{,t} - {\mathcal C}_{2211} v^5_{,2}=0,
\end{equation}
is used. Adding the last three equations we get the new equation used
\begin{eqnarray}
&& {\mathcal C}_{1111} v^1_{,t}+2{\mathcal C}_{2111} v^2_{,t}+{\mathcal C}_{2211} v^4_{,t} \nonumber\\
&& -{\mathcal C}_{1111} v^3_{,1}-{\mathcal C}_{2111} v^3_{,2}-{\mathcal C}_{2111} v^5_{,1}-{\mathcal C}_{2211} v^5_{,2}=0
\end{eqnarray}
which changes the first line of the matrix  $A^{({\bf u}_t, {\bf e})}$ as
\begin{equation}
A^{({\bf u}_t, {\bf e})}=\begin{pmatrix}
{\mathcal C}_{1111} & 2{\mathcal C}_{2111}  & 0 & {\mathcal C}_{2211} & 0 \\
0 & 2  & 0 & 0 & 0 \\
0 & 0 & 1 & 0 & 0 \\
0 & 0 & 0 & 1 & 0 \\
0 & 0 & 0 & 0 & 1 
\end{pmatrix}.
\end{equation}

For the second line the initial equation used is
\begin{equation}
2v^2_{,t}-v^3_{,2}-v^5_{,1}=0. 
\end{equation}
In order to tackle the (2, 5)-element of $B_{1 \ \  \text{sym}}^{({\bf u}_t, {\bf e})}$ we multiply the last equation with $2 {\mathcal C}_{2112}$ to get
\begin{equation}
4 {\mathcal C}_{2112} v^2_{,t}-2 {\mathcal C}_{2112} v^3_{,2}-2 {\mathcal C}_{2112} v^5_{,1}=0, 
\end{equation}
which takes care also the (2, 3)-element of $B_{2 \ \  \text{sym}}^{({\bf u}_t, {\bf e})}$  which has the elastic constant ${\mathcal C}_{1212}$, but the minor symmetry helps to reconcile it. For the (2, 3)-element of $B_{1 \ \  \text{sym}}^{({\bf u}_t, {\bf e})}$ we work with 
\begin{equation}
2 {\mathcal C}_{1112} v^1_{,t}-2 {\mathcal C}_{1112} v^3_{,1}=0, 
\end{equation}
while for the (2, 5)-element of $B_{2 \ \  \text{sym}}^{({\bf u}_t, {\bf e})}$ we use
\begin{equation}
2 {\mathcal C}_{2212} v^4_{,t}-2 {\mathcal C}_{2212} v^5_{,2}=0.
\end{equation}
Adding the last three equations we get the new eqution that should be used
\begin{eqnarray}
&& 2 {\mathcal C}_{1112} v^1_{,t}+4{\mathcal C}_{2112} v^2_{,t}+2{\mathcal C}_{2212} v^4_{,t} \nonumber\\
&& -2{\mathcal C}_{1112} v^3_{,1}-2{\mathcal C}_{2112} v^3_{,2}-2{\mathcal C}_{2112} v^5_{,1}-2{\mathcal C}_{2212} v^5_{,2}=0.
\end{eqnarray}
When this last equation is used the matrix $A^{({\bf u}_t, {\bf e})}$ has the form
\begin{equation}
A^{({\bf u}_t, {\bf e})}=\begin{pmatrix}
{\mathcal C}_{1111} & 2{\mathcal C}_{2111}  & 0 & {\mathcal C}_{2211} & 0 \\
2 {\mathcal C}_{1112} & 4{\mathcal C}_{2112} & 0 & 2{\mathcal C}_{2212} & 0 \\
0 & 0 & 1 & 0 & 0 \\
0 & 0 & 0 & 1 & 0 \\
0 & 0 & 0 & 0 & 1 
\end{pmatrix},
\end{equation}
The third and fifth line do not change since these are the momentum equations. 

For the fourth line the starting point is the compatibility equation
\begin{equation}
v^2_{,t}-v^5_{,2}=0, 
\end{equation}
which is multiplied with ${\mathcal C}_{2222}$ in order to built the (4, 5)-element of $B_{2 \ \  \text{sym}}^{({\bf u}_t, {\bf e})}$ 
\begin{equation}
{\mathcal C}_{2222} v^2_{,t} - {\mathcal C}_{2222} v^5_{,2}=0, 
\end{equation}
For the  (4, 5)-element of $B_{1 \ \  \text{sym}}^{({\bf u}_t, {\bf e})}$ and the (4, 3)-element of $B_{2 \ \  \text{sym}}^{({\bf u}_t, {\bf e})}$ we use
\begin{equation}
2 {\mathcal C}_{1222} v^2_{,t} - {\mathcal C}_{1222} v^3_{,2} - {\mathcal C}_{2122} v^5_{,1} =0, 
\end{equation}
since again the minor symmetries here play a crucial role. What it remains is the (4, 3)-element of $B_{1 \ \  \text{sym}}^{({\bf u}_t, {\bf e})}$ for which we have
\begin{equation}
{\mathcal C}_{1122} v^1_{,t} - {\mathcal C}_{1122} v^3_{,1}=0. 
\end{equation}
Adding the last three equations we have
\begin{eqnarray}
&&  {\mathcal C}_{1122} v^1_{,t}+2{\mathcal C}_{1222} v^2_{,t}+{\mathcal C}_{2222} v^4_{,t} \nonumber\\
&& -{\mathcal C}_{1122} v^3_{,1}-{\mathcal C}_{1222} v^3_{,2}-{\mathcal C}_{2122} v^5_{,1}-{\mathcal C}_{2222} v^5_{,2}=0.
\end{eqnarray}
When this last equation is used the matrix $A^{({\bf u}_t, {\bf e})}$ has the form
\begin{equation}
A_{\text{sym}}^{({\bf u}_t, {\bf e})}=\begin{pmatrix}
{\mathcal C}_{1111} & 2{\mathcal C}_{2111}  & 0 & {\mathcal C}_{2211} & 0 \\
2 {\mathcal C}_{1112} & 4{\mathcal C}_{2112} & 0 & 2{\mathcal C}_{2212} & 0 \\
0 & 0 & 1 & 0 & 0 \\
{\mathcal C}_{1122} & 2 {\mathcal C}_{1222} & 0 & {\mathcal C}_{2222} & 0 \\
0 & 0 & 0 & 0 & 1 
\end{pmatrix},
\end{equation}
which is obviously symmetric when major and minor symmetries are enforced. So, all in all, we built a symmetric writing of the form of eq. (73) with $A_{\text{sym}}^{({\bf u}_t, {\bf e})}$, $B_{1 \text{sym}}^{({\bf u}_t, {\bf e})}$ and $B_{2 \text{sym}}^{({\bf u}_t, {\bf e})}$ given by eqs. (112, 93, 94). The determinant of the matrix $A_{\text{sym}}^{({\bf u}_t, {\bf e})}$ is non-negative since
\begin{equation}
	\text{det} A_{\text{sym}}^{({\bf u}_t, {\bf e})}=\text{det}\begin{pmatrix}
		{\mathcal C}_{1111} & 2{\mathcal C}_{2111}  & {\mathcal C}_{2211}  \\
		2 {\mathcal C}_{1112} & 4{\mathcal C}_{2112} &  2{\mathcal C}_{2212}  \\
		{\mathcal C}_{1122} & 2 {\mathcal C}_{1222} & {\mathcal C}_{2222} \\
	\end{pmatrix} \neq 0.
\end{equation} 
 
\underline{Remark} \\
Compared to the writing given for the two dimensional case in \cite{Sfyris2024}, we here have 5 equations while there we had 6. This is due to a mistake made in \cite{Sfyris2024}. While eq. (44) of \cite{Sfyris2024} is equivalent to the invertibility of the matrix of elasticities, the compatibility equations used there are not independent. More specifically, the second second and fourth line of eq. (40) of \cite{Sfyris2024} for the matrix $A^{2d}$ are the same when major and minor symmetries are enforced. So, at the end of the day, the matrix $A^{2d}$ is positive semi-definite, and a similar remark holds for the three dimensional case presented there. The writing presented there remains symmetric nevertheless and might help when comparing it with its nonlinear counterpart (\cite{Boillat-Ruggeri}).

\underline{Remark} \\
An alternative writing of the equations of linear elastodynamics using the strain components is given by \cite{Morando-Serre}. This writing is valid for the isotropic case and the main difference with our approach here is that the matrix multiplying the time derivatives is a unit matrix in \cite{Morando-Serre}. 
 
\subsection{Elasticity in 2D with respect to $({\bf u}_t, {\boldsymbol \sigma})$}

We follow the approach of Yakhno-Akmaz (\cite{Yakhno-Akmaz}) who use the momentum equation and the constitutive law time differentiated. From the momentum equation we have $\ddot{u}_i=\sigma_{ij,j}$ which reads
\begin{eqnarray}
\ddot{u}_1&=&\sigma_{11,1}+\sigma_{12,2}, \\
\ddot{u}_2&=&\sigma_{21,1}+\sigma_{22,2}.
\end{eqnarray} 
For the unknowns we set
\begin{eqnarray}
{\bf q}=&&\{v^1=\dot{u}_1, v^2=\dot{u}_2, \nonumber\\ 
&& v^3=\sigma_{11}, v^4=\sigma_{22}, v^5=\sigma_{12}=\sigma_{21}  \},
\end{eqnarray}
so that the momentum equations read
\begin{eqnarray}
v^1_{,t}-v^3_{,1}-v^5_{,2}&=&0, \\
v^2_{,t}-v^5_{,1}-v^4_{,2}&=&0.
\end{eqnarray}

The rest three equations that should be used are the constitutive law time differentiated. Which means that we work with 
\begin{equation}
\sigma_{ij}=\mathcal C_{ijkl} e_{kl} \rightarrow {\mathcal C}^{-1}_{ijkl} \dot{\sigma}_{kl}=\dot{e}_{kl}.
\end{equation}
Now, the relation between strains and displacement gradients, eq. (67), time differentiated renders
\begin{equation}
\dot{e}_{ij}=\frac{1}{2}(\dot{u}_{i,j}+\dot{u}_{j,i}),
\end{equation}
which reads
\begin{eqnarray}
\dot{e}_{11}&=&\dot{u}_{1,1}=v^1_{,1}, \\
\dot{e}_{22}&=&\dot{u}_{2,2}=v^2_{,2}, \\
\dot{e}_{12}&=&\frac{1}{2}(\dot{u}_{1,2}+\dot{u}_{2,1})=\frac{1}{2}(v^1_{,2}+v^2_{,1}) \rightarrow 2\dot{e}_{12}=v^1_{,2}+v^2_{,1}.
\end{eqnarray} 

The time differentiated constitutive laws, eqs. (119) are
\begin{eqnarray}
{\mathcal C}^{-1}_{1111} \dot{\sigma}_{11} + {\mathcal C}^{-1}_{1112} \dot{\sigma}_{12} + {\mathcal C}^{-1}_{1121} \dot{\sigma}_{21}+{\mathcal C}^{-1}_{1122} \dot{\sigma}_{22}&=&\dot{e}_{11}, \\
{\mathcal C}^{-1}_{2211} \dot{\sigma}_{11} + {\mathcal C}^{-1}_{2212} \dot{\sigma}_{12} + {\mathcal C}^{-1}_{2221} \dot{\sigma}_{21}+{\mathcal C}^{-1}_{2222} \dot{\sigma}_{22}&=&\dot{e}_{22}, \\
{\mathcal C}^{-1}_{1211} \dot{\sigma}_{11} + {\mathcal C}^{-1}_{1212} \dot{\sigma}_{12} + {\mathcal C}^{-1}_{1221} \dot{\sigma}_{21}+{\mathcal C}^{-1}_{1222} \dot{\sigma}_{22}&=&\dot{e}_{12}.
\end{eqnarray} 
In terms of the new variables these last three equations read
\begin{eqnarray}
{\mathcal C}^{-1}_{1111} v^3_{,t}+({\mathcal C}^{-1}_{1112}+{\mathcal C}^{-1}_{1121})v^5_{,t}+{\mathcal C}^{-1}_{1122}v^4_{,2}=v^1_{,1} \\
{\mathcal C}^{-1}_{2211} v^3_{,t}+({\mathcal C}^{-1}_{2212}+{\mathcal C}^{-1}_{2221})v^5_{,t}+{\mathcal C}^{-1}_{2222}v^4_{,2}=v^2_{,2} \\
2({\mathcal C}^{-1}_{1211} v^3_{,t}+({\mathcal C}^{-1}_{1212}+{\mathcal C}^{-1}_{1221})v^5_{,t}+{\mathcal C}^{-1}_{1222}v^4_{,2})=v^1_{,2}+v^2_{,1}.
\end{eqnarray} 

So, when eqs (117, 118, 127, 128, 129) are written in the form
\begin{equation}
A_{\text{sym}}^{({\bf u}_t, {\boldsymbol \sigma})} \frac{\partial {\bf q}}{\partial t} + \sum_{i=1}^2 B_{i \ \text{sym}}^{({\bf u}_t, {\boldsymbol \sigma})} \frac{\partial {\bf q}}{\partial x_i}={\bf 0},
\end{equation} 
 with $\bf q$ from eq. (116) and
\begin{equation}
A_{\text{sym}}^{({\bf u}_t, {\boldsymbol \sigma})}=\begin{pmatrix}
1 & 0  & 0 & 0 & 0 \\
0 & 1  & 0 & 0 & 0 \\
0 & 0 & {\mathcal C}^{-1}_{1111} & {\mathcal C}^{-1}_{1122}  & 2{\mathcal C}^{-1}_{1112} \\
0 & 0 & {\mathcal C}^{-1}_{2211} & {\mathcal C}^{-1}_{2222}  & 2{\mathcal C}^{-1}_{2212} \\
0 & 0 & 2{\mathcal C}^{-1}_{1211} & 2{\mathcal C}^{-1}_{1222}  & 4{\mathcal C}^{-1}_{1212} 
\end{pmatrix},
\end{equation}
and
\begin{equation}
B_{1 \ \text{sym}}^{({\bf u}_t, {\boldsymbol \sigma})}=\begin{pmatrix}
0 & 0  & -1 & 0 & 0 \\
0 & 0  & 0 & 0 & -1 \\
-1 & 0 & 0 & 0 & 0 \\
0 & 0 & 0 & 0 & 0 \\
0 & -1 & 0 & 0 & 0
\end{pmatrix},
\end{equation}
as well as 
\begin{equation}
B_{2 \ \text{sym}}^{({\bf u}_t, {\boldsymbol \sigma})}=\begin{pmatrix}
0 & 0  & 0 & 0 & -1 \\
0 & 0  & 0 & -1 & 0 \\
0 & 0 & 0 & 0 & 0 \\
0 & -1 & 0 & 0 & 0 \\
-1 & 0 & 0 & 0 & 0 
\end{pmatrix},
\end{equation}
it is clearly in a symmetric format when major and minor symmetries are enforced. 

\subsection{Elasticity in 2D with respect to $({\bf u}_t, {\bf e}, {\boldsymbol \sigma})$}

In the one dimensional setting in order to get a symmetric writing with respect to all three variables we added the momentum equation, once written with respect to $({\bf u}_t, {\bf e})$ and once with respect to $({\bf u}_t, {\boldsymbol \sigma})$; we here examine if this route lead us to a symmetric writing again. The starting point is to choose the new variables. We make the choice
\begin{eqnarray}
{\bf q}=&&\{v^1=e_{11}=u_{1,1}, v^2=\frac{1}{2}(u_{1,2}+u_{2,1})=e_{12}, v^3=u_{1,t},, v^4=e_{22}=u_{2,2}, \nonumber\\ 
&& v^5=u_{2,t}, v^6=\sigma_{11}, v^7=\sigma_{22}, v^8=\sigma_{12}=\sigma_{21}  \},
\end{eqnarray}
and the equations which we use are the momentum equation, the compatibility and the constitutive law time differentiated. First, we should say that we need eight equations. For the first five we use the same as that in the writing using  $({\bf u}_t, {\bf e})$ and to them we add the constitutive law time differentiated. 

So, with respect to the new variables the first equation used is
\begin{eqnarray}
&& {\mathcal C}_{1111} v^1_{,t}+2{\mathcal C}_{2111} v^2_{,t}+{\mathcal C}_{2211} v^4_{,t} \nonumber\\
&& -{\mathcal C}_{1111} v^3_{,1}-{\mathcal C}_{2111} v^5_{,1}-{\mathcal C}_{1211} v^3_{,2}-{\mathcal C}_{2211} v^5_{,2}=0,
\end{eqnarray}
which is a compatibility equation. The second equation used, which again is a compatibility equation reads
\begin{eqnarray}
&& 2{\mathcal C}_{1112} v^1_{,t}+4{\mathcal C}_{2112} v^2_{,t}+2{\mathcal C}_{2212} v^4_{,t} \nonumber\\
&& -2{\mathcal C}_{1112} v^3_{,1}-2{\mathcal C}_{2112} v^5_{,1}-2{\mathcal C}_{1212} v^3_{,2}-{\mathcal C}_{2212} v^5_{,2}=0.
\end{eqnarray}
The third equation used is the momentum equation which reads
\begin{eqnarray}
&& v^3_{,t}-{\mathcal C}_{1111} v^1_{,1}-2{\mathcal C}_{1112} v^2_{,1}-{\mathcal C}_{1122} v^4_{,1} \nonumber\\
&& -{\mathcal C}_{1211} v^1_{,2}-2{\mathcal C}_{1212} v^2_{,2}-{\mathcal C}_{1222} v^4_{,2}=0.
\end{eqnarray}
The fourth equation is a compatibility equation and reads
\begin{eqnarray}
&& {\mathcal C}_{1122} v^1_{,t}+2{\mathcal C}_{1222} v^2_{,t}+{\mathcal C}_{2222} v^4_{,t} \nonumber\\
&& -{\mathcal C}_{1122} v^3_{,1}-{\mathcal C}_{2122} v^5_{,1}-{\mathcal C}_{1222} v^3_{,2}-{\mathcal C}_{2222} v^5_{,2}=0,
\end{eqnarray}
and the fifth equation which is the momentum equation has the form
\begin{eqnarray}
&& v^5_{,t}-{\mathcal C}_{2111} v^1_{,1}-2{\mathcal C}_{2112} v^2_{,1}-{\mathcal C}_{2122} v^4_{,1} \nonumber\\
&& -{\mathcal C}_{2211} v^1_{,2}-2{\mathcal C}_{2212} v^2_{,2}-{\mathcal C}_{2222} v^4_{,2}=0.
\end{eqnarray}

Now, we should use the constitutive laws time differentiated, but written with respect to the new variables. The sixth equation of the system is thus
\begin{equation}
{\mathcal C}^{-1}_{1111} v^6_{,t}+{\mathcal C}^{-1}_{1122} v^7_{,t}+2{\mathcal C}^{-1}_{1112} v^8_{,t}-v^3_{,1}=0, 
\end{equation}
the seventh equation reads
\begin{equation}
{\mathcal C}^{-1}_{2211} v^6_{,t}+2 {\mathcal C}^{-1}_{2212} v^7_{,t}+{\mathcal C}^{-1}_{2222} v^8_{,t}-v^5_{,2}=0, 
\end{equation}
while the eighth equation is 
\begin{equation}
2{\mathcal C}^{-1}_{1211} v^6_{,t}+2 {\mathcal C}^{-1}_{1222} v^7_{,t}+4{\mathcal C}^{-1}_{1212} v^8_{,t}-v^3_{,2}-v^5_{,1}=0. 
\end{equation}

So, as a system of the form 
\begin{equation}
A^{({\bf u}_t, {\bf e}, {\boldsymbol \sigma})} \frac{\partial {\bf q}}{\partial t} + \sum_{i=1}^2 B_i^{({\bf u}_t, {\bf e}, {\boldsymbol \sigma})} \frac{\partial {\bf q}}{\partial x_i}={\bf 0},
\end{equation} 
with $\bf q$ from eq. (134) and
\begin{equation}
A^{({\bf u}_t, {\bf e}, {\boldsymbol \sigma})}=\begin{pmatrix}
{\mathcal C}_{1111} & 2{\mathcal C}_{2111}  & 0 & {\mathcal C}_{2211} & 0 & 0 & 0 & 0 \\
2{\mathcal C}_{1112} & 4{\mathcal C}_{2112}  & 0 & 2{\mathcal C}_{2212} & 0 & 0 & 0 & 0 \\
0 & 0 & 1 & 0 & 0 & 0 & 0 & 0 \\
{\mathcal C}_{1122} & 2 {\mathcal C}_{1222} & 0 & {\mathcal C}_{2222} & 0 & 0 & 0 & 0 \\
0 & 0 & 0 & 0 & 1 & 0 & 0 & 0 \\ 
0 & 0 & 0 & 0 & 0 & {\mathcal C}^{-1}_{1111} & {\mathcal C}^{-1}_{1122}  & 2{\mathcal C}^{-1}_{1112} \\
0 & 0 & 0 & 0 & 0 & {\mathcal C}^{-1}_{2211} & {\mathcal C}^{-1}_{2222}  & 2{\mathcal C}^{-1}_{2212} \\
0 & 0 & 0 & 0 & 0 & 2{\mathcal C}^{-1}_{1211} & 2{\mathcal C}^{-1}_{1222}  & 4{\mathcal C}^{-1}_{1212} 
\end{pmatrix},
\end{equation}
which is clearly symmetric. We also have 
\begin{equation}
B_1^{({\bf u}_t, {\bf e}, {\boldsymbol \sigma})}=\begin{pmatrix}
0 & 0  & -{\mathcal C}_{1111} & 0 & -{\mathcal C}_{2111} & 0 & 0 & 0 \\
0 & 0  & -2{\mathcal C}_{1112} & 0 & -2{\mathcal C}_{2112} & 0 & 0 & 0 \\
-{\mathcal C}_{1111} & -2{\mathcal C}_{1112} & 0 & -{\mathcal C}_{1122} & 0 & 0 & 0 & 0\\
0 & 0 & -{\mathcal C}_{1122} & 0 & -{\mathcal C}_{2122} & 0 & 0 & 0\\
-{\mathcal C}_{2111} & -2{\mathcal C}_{2112} & 0 & -{\mathcal C}_{2122} & 0 & 0 & 0 & 0 \\
0 & 0 & -1 & 0 & 0 & 0 & 0 & 0 \\
0 & 0 & 0 & 0 & 0 & 0 & 0 & 0 \\
0 & 0 & 0 & 0 & -1 & 0 & 0 & 0 
\end{pmatrix},
\end{equation}
which is clearly not symmetric due to terms -1 in the (6, 3) and (8, 5) element of the matrix. In order to make this matrix symmetric we have to add in the (3, 6) and (5, 8) elements the term -1. This means that we have to add to the third equation the term $-v^6_{1}=\sigma_{11,1}$ and to the fifth equation the term $-v^8_{,1}=-\sigma_{21,1}$. For the other matrix we have 
\begin{equation}
B_2^{({\bf u}_t, {\bf e}, {\boldsymbol \sigma})}=\begin{pmatrix}
0 & 0  & -{\mathcal C}_{1211} & 0 & -{\mathcal C}_{2211} & 0 & 0 & 0 \\
0 & 0  & -2{\mathcal C}_{1212} & 0 & -2{\mathcal C}_{2212} & 0 & 0 & 0 \\
-{\mathcal C}_{1211} & -2{\mathcal C}_{1212} & 0 & -{\mathcal C}_{1222} & 0 & 0 & 0 & 0\\
0 & 0 & -{\mathcal C}_{1222} & 0 & -{\mathcal C}_{2222} & 0 & 0 & 0\\
-{\mathcal C}_{2211} & -2{\mathcal C}_{2212} & 0 & -{\mathcal C}_{2222} & 0 & 0 & 0 & 0 \\
0 & 0 & 0 & 0 & 0 & 0 & 0 & 0 \\
0 & 0 & 0 & 0 & -1 & 0 & 0 & 0 \\
0 & 0 & -1 & 0 & 0 & 0 & 0 & 0 
\end{pmatrix},
\end{equation}
which is not symmetric as well due to the terms -1 in the (7, 5) and (8, 3) element of the matrix. In order to make this matrix symmetric we have to add in the (5, 7) and (3, 8) elements the terms -1. Namely, we have to add to the third equation the term $-v^8_{,2}=-\sigma_{21,2}$ and in the fifth equation the term $-v^7_{,2}=-\sigma_{22,2}$. 

So, now, we collect which terms should be added in order to make the matrices $B_i^{({\bf u}_t, {\bf e}, {\boldsymbol \sigma})}, i=1, 2$ symmetric. In the third equation we have to add terms $-v^6_{,1}-v^8_{,2}=-\sigma_{11,1}-\sigma_{12,2}$. In the fifth equation we have to add terms $-v^8_{,1}-v^7_{,2}=-\sigma_{21,1}-\sigma_{22,2}$. On the other hand the momentum equation written with respect to the stresses is 
\begin{eqnarray}
&& v^3_{,t}-v^6_{,1}-v^8_{,2}=0, \\
&& v^5_{,t}-v^8_{,1}-v^7_{,2}=0. 
\end{eqnarray}
 So, all in all, if we add this expression of the momentum equation in the third and fifth equation of the system (which again are the momentum equations) this procedure render the matrices $B_i^{({\bf u}_t, {\bf e}, {\boldsymbol \sigma})}, i=1, 2$ symmetric. What changes in the $A^{({\bf u}_t, {\bf e}, {\boldsymbol \sigma})}$ matrix is that we have 2 instead of 1 in the (3, 3) and (5, 5) elements. So, collectively this symmetric writing has the following matrices
\begin{equation}
A_{\text{sym}}^{({\bf u}_t, {\bf e}, {\boldsymbol \sigma})}=\begin{pmatrix}
{\mathcal C}_{1111} & 2{\mathcal C}_{2111}  & 0 & {\mathcal C}_{2211} & 0 & 0 & 0 & 0 \\
2{\mathcal C}_{1112} & 4{\mathcal C}_{2112}  & 0 & 2{\mathcal C}_{2212} & 0 & 0 & 0 & 0 \\
0 & 0 & 2 & 0 & 0 & 0 & 0 & 0 \\
{\mathcal C}_{1122} & 2 {\mathcal C}_{1222} & 0 & {\mathcal C}_{2222} & 0 & 0 & 0 & 0 \\
0 & 0 & 0 & 0 & 2 & 0 & 0 & 0 \\ 
0 & 0 & 0 & 0 & 0 & {\mathcal C}^{-1}_{1111} & {\mathcal C}^{-1}_{1122}  & 2{\mathcal C}^{-1}_{1112} \\
0 & 0 & 0 & 0 & 0 & {\mathcal C}^{-1}_{2211} & {\mathcal C}^{-1}_{2222}  & 2{\mathcal C}^{-1}_{2212} \\
0 & 0 & 0 & 0 & 0 & 2{\mathcal C}^{-1}_{1211} & 2{\mathcal C}^{-1}_{1222}  & 4{\mathcal C}^{-1}_{1212} 
\end{pmatrix},
\end{equation}
and
\begin{equation}
B_{1 \ \text{sym}}^{({\bf u}_t, {\bf e}, {\boldsymbol \sigma})}=\begin{pmatrix}
0 & 0  & -{\mathcal C}_{1111} & 0 & -{\mathcal C}_{2111} & 0 & 0 & 0 \\
0 & 0  & -2{\mathcal C}_{1112} & 0 & -2{\mathcal C}_{2112} & 0 & 0 & 0 \\
-{\mathcal C}_{1111} & -2{\mathcal C}_{1112} & 0 & -{\mathcal C}_{1122} & 0 & -1 & 0 & 0\\
0 & 0 & -{\mathcal C}_{1122} & 0 & -{\mathcal C}_{2122} & 0 & 0 & 0\\
-{\mathcal C}_{2111} & -2{\mathcal C}_{2112} & 0 & -{\mathcal C}_{2122} & 0 & 0 & 0 & -1 \\
0 & 0 & -1 & 0 & 0 & 0 & 0 & 0 \\
0 & 0 & 0 & 0 & 0 & 0 & 0 & 0 \\
0 & 0 & 0 & 0 & -1 & 0 & 0 & 0 
\end{pmatrix},
\end{equation}
and
\begin{equation}
B_{2 \ \text{sym}}^{({\bf u}_t, {\bf e}, {\boldsymbol \sigma})}=\begin{pmatrix}
0 & 0  & -{\mathcal C}_{1211} & 0 & -{\mathcal C}_{2211} & 0 & 0 & 0 \\
0 & 0  & -2{\mathcal C}_{1212} & 0 & -2{\mathcal C}_{2212} & 0 & 0 & 0 \\
-{\mathcal C}_{1211} & -2{\mathcal C}_{1212} & 0 & -{\mathcal C}_{1222} & 0 & 0 & 0 & -1 \\
0 & 0 & -{\mathcal C}_{1222} & 0 & -{\mathcal C}_{2222} & 0 & 0 & 0\\
-{\mathcal C}_{2211} & -2{\mathcal C}_{2212} & 0 & -{\mathcal C}_{2222} & 0 & 0 & -1 & 0 \\
0 & 0 & 0 & 0 & 0 & 0 & 0 & 0 \\
0 & 0 & 0 & 0 & -1 & 0 & 0 & 0 \\
0 & 0 & -1 & 0 & 0 & 0 & 0 & 0 
\end{pmatrix}.
\end{equation}
This approach renders the symmetric writing we seek for. 

\subsection{Elasticity in 2D with respect to $(\rho_0 {\bf u}_t, {\bf u}_t, {\bf e}, {\boldsymbol \sigma})$}

In the one dimensional case in order to arrive at a symmetric format using the four variables we added the momentum equations in order to have the quantities $(u_t,\sigma)$ and $(\rho_0 u_t, \sigma)$ and used these momentum equation twice as a field equation. These are the first two equations in the one dimensional case. The next equation is the compatibility equation for which we add the compatibility equation written once with respect to $\rho_0  u_t$ and once with respect to $u_t$. What it remains is the constitutive law time differentiated; for this we also need a double writing: once with respect to $\rho_0 u_t$ and once with respect to $u_t$.

In order to avoid cumbersome calculations we  essentially reiterate the approach of the one dimensional case, in order not to come back and symmetrize each matrix. 

The equations are 10 in total. The sequence of the equations is: the first two equations are the momentum equations written with respect to $\rho_0 {\bf u}_t$ and the strain tensor added again to the momentum equation written with respect to $\rho_0 {\bf u}_t$ but using the stress tensor instead of the strain tensor. Then we follow the approach of the writing using $({\bf u}_t, {\bf e}, {\boldsymbol \sigma})$. This means that the third, fourth and sixth are compatibility equations and the fifth and seventh are momentum equations. The three last equations are the time differentiated constitutive laws. But we stress that for the momentum equation (fifth and seventh equation) we add the momentum equation once with respect to $({\bf u}_t, {\bf e})$ and once with respect to $({\bf u}_t, {\boldsymbol \sigma})$. The compatibility equations is the addition of the corresponding compatibility equations once written with respect to $\rho_0 {\bf u}_t$ and once with respect to ${\bf u}_t$. The same is done for the constitutive laws: we add the constitutive laws written once with respect to $\rho_0 {\bf u}_t$ and once with respect to ${\bf u}_t$.

For the variables used we set
\begin{eqnarray}
	{\bf q}=&&\{v^1=\rho_0 u_{1,t}, v^2=\rho_0 u_{2,t}, v^3=e_{11}=u_{1,1}, v^4=\frac{1}{2}(u_{1,2}+u_{2,1})=e_{12}, \\
	&& v^5=u_{1,t}, v^6=e_{22}=u_{2,2}, v^7=u_{2,t}, v^8=\sigma_{11}, v^9=\sigma_{22}, v^{10}=\sigma_{12}=\sigma_{21}  \} \nonumber,
\end{eqnarray}
The first of the momentum equation written with respect to the strain components in the new system of variables read
\begin{eqnarray}
	v^1_{,t}&&{-\mathcal C}_{1111} v^3_{,1}-{\mathcal C}_{1211} v^3_{,2}-{\mathcal C}_{1112} 2 v^4_{,1} \nonumber\\
	&&-{\mathcal C}_{1212} 2 v^4_{,2}-{\mathcal C}_{1122} v^6_{,1}-{\mathcal C}_{1222} v^6_{,2}=0,
\end{eqnarray}
and on the other hand the same momentum equation written with respect to stresses reads
\begin{equation}
	v^1_{,t}-v^8_{,1}-v^{10}_{,2}=0.
\end{equation}
We should now add the last two equations in order to give us the first equation of our new writing:
\begin{eqnarray}
	2v^1_{,t}&&{-\mathcal C}_{1111} v^3_{,1}-{\mathcal C}_{1211} v^3_{,2}-{\mathcal C}_{1112} 2 v^4_{,1} \nonumber\\
	&&-{\mathcal C}_{1212} 2 v^4_{,2}-{\mathcal C}_{1122} v^6_{,1}-{\mathcal C}_{1222} v^6_{,2} \nonumber\\
	&&-v^8_{,1}-v^{10}_{,2}=0
\end{eqnarray}

For the second of the momentum equation we do exactly the same thing: once written with respect to the strains
\begin{eqnarray}
	v^2_{,t}&&{-\mathcal C}_{2111} v^3_{,1}-{\mathcal C}_{2211} v^3_{,2}-{\mathcal C}_{2112} 2 v^4_{,1} \nonumber\\
	&&-{\mathcal C}_{2212} 2 v^4_{,2}-{\mathcal C}_{2122} v^6_{,1}-{\mathcal C}_{2222} v^6_{,2}=0,
\end{eqnarray}
while the momentum equation written with respect to the stresses reads
\begin{equation}
	v^2_{,t}-v^{10}_{,1}-v^9_{,2}=0.
\end{equation}
If we add these two last equations we get the second equation of our system 
\begin{eqnarray}
	v^2_{,t}&&{-\mathcal C}_{2111} v^3_{,1}-{\mathcal C}_{2211} v^3_{,2}-{\mathcal C}_{2112} 2 v^4_{,1} \nonumber\\
	&&-{\mathcal C}_{2212} 2 v^4_{,2}-{\mathcal C}_{2122} v^6_{,1}-{\mathcal C}_{2222} v^6_{,2} \nonumber\\
	&&-v^{10}_{,1}-v^9_{,2}=0.
\end{eqnarray}

For the rest equations we follow the approach of the $({\bf u}_t, {\bf e}, {\boldsymbol \sigma})$ writing; the third, fourth and sixth are compatibility equations, the fifth and seventh are momentum and the last three equations are constitutive laws time differentiated. For the compatibility equations we add the corresponding compatibility equation once written with respect to $\rho_0 {\bf u}_t$ and once written with respect to ${\bf u}_t$. 

So, for the third equation we multiply the corresponding compatibility equation by $\rho_0$ and directly obtain 
\begin{eqnarray}
	&& \rho_0 {\mathcal C}_{1111} v^3_{,t} + 2 \rho_0 {\mathcal C}_{2111} v^4_{,t} + \rho_0 {\mathcal C}_{2211} v^6_{,t} \nonumber\\
	&&-{\mathcal C}_{1111} v^1_{,1} - {\mathcal C}_{2111} v^2_{,1} - {\mathcal C}_{1211} v^1_{,2} - {\mathcal C}_{2211} v^2_{,2}=0,
\end{eqnarray} 
which should be added to 
\begin{eqnarray}
	&& {\mathcal C}_{1111} v^3_{,t} + 2 {\mathcal C}_{2111} v^4_{,t} + {\mathcal C}_{2211} v^6_{,t} \nonumber\\
	&&-{\mathcal C}_{1111} v^5_{,1} - {\mathcal C}_{2111} v^7_{,1} - {\mathcal C}_{1211} v^5_{,2} - {\mathcal C}_{2211} v^7_{,2}=0,
\end{eqnarray} 
in order to, collectively, take
\begin{eqnarray}
	&& (\rho_0+1) {\mathcal C}_{1111} v^3_{,t} + 2 (\rho_0+1) {\mathcal C}_{2111} v^4_{,t} + (\rho_0+1) {\mathcal C}_{2211} v^6_{,t} \nonumber\\
	&&-{\mathcal C}_{1111} v^1_{,1} - {\mathcal C}_{2111} v^2_{,1} - {\mathcal C}_{1211} v^1_{,2} - {\mathcal C}_{2211} v^2_{,2} \nonumber\\
	&&-{\mathcal C}_{1111} v^5_{,1} - {\mathcal C}_{2111} v^7_{,1} - {\mathcal C}_{1211} v^5_{,2} - {\mathcal C}_{2211} v^7_{,2}=0.
\end{eqnarray} 

For the fourth equation of our system we have when written with respect to $\rho_0 {\bf u}_t$
\begin{eqnarray}
	&& 2 \rho_0 {\mathcal C}_{1112} v^3_{,t} + 4 \rho_0 {\mathcal C}_{2112} v^4_{,t} + 2 \rho_0 {\mathcal C}_{2212} v^6_{,t} \nonumber\\
	&&- 2 {\mathcal C}_{1112} v^1_{,1} - 2 {\mathcal C}_{2112} v^2_{,1} - 2 {\mathcal C}_{1212} v^1_{,2} - 2 {\mathcal C}_{2212} v^2_{,2}=0,
\end{eqnarray} 
which should be added to 
\begin{eqnarray}
	&& 2 {\mathcal C}_{1112} v^3_{,t} + 4 {\mathcal C}_{2112} v^4_{,t} + 2 {\mathcal C}_{2212} v^6_{,t} \nonumber\\
	&&- 2 {\mathcal C}_{1112} v^5_{,1} - 2 {\mathcal C}_{2112} v^7_{,1} - 2 {\mathcal C}_{1212} v^5_{,2} - 2 {\mathcal C}_{2212} v^7_{,2}=0,
\end{eqnarray} 
to finally give 
\begin{eqnarray}
	&& 2 (\rho_0+1) {\mathcal C}_{1112} v^3_{,t} + 4 (\rho_0+1) {\mathcal C}_{2112} v^4_{,t} + 2 (\rho_0+1) {\mathcal C}_{2212} v^6_{,t} \nonumber\\
	&&- 2 {\mathcal C}_{1112} v^1_{,1} - 2 {\mathcal C}_{2112} v^2_{,1} - 2 {\mathcal C}_{1212} v^1_{,2} - 2 {\mathcal C}_{2212} v^2_{,2} \nonumber\\
	&& - 2 {\mathcal C}_{1112} v^5_{,1} - 2 {\mathcal C}_{2112} v^7_{,1} - 2 {\mathcal C}_{1212} v^5_{,2} - 2 {\mathcal C}_{2212} v^7_{,2}=0.
\end{eqnarray} 

The fifth equation is the first of the momentum equations written with respect to ${\bf u}_t$
\begin{eqnarray}
	\rho_0 v^5_{,t}&&{-\mathcal C}_{1111} v^3_{,1}-{\mathcal C}_{1211} v^3_{,2}-{\mathcal C}_{1112} 2 v^4_{,1} \nonumber\\
	&&-{\mathcal C}_{1212} 2 v^4_{,2}-{\mathcal C}_{1122} v^6_{,1}-{\mathcal C}_{1222} v^6_{,2}=0,
\end{eqnarray}
and on the other hand the same momentum equation written with respect to stresses reads
\begin{equation}
	\rho_0 v^5_{,t}-v^8_{,1}-v^{10}_{,2}=0.
\end{equation}
We should now add the last two equations in order to give us the first equation of our new writing:
\begin{eqnarray}
	2 \rho_0 v^5_{,t}&&{-\mathcal C}_{1111} v^3_{,1}-{\mathcal C}_{1211} v^3_{,2}-{\mathcal C}_{1112} 2 v^4_{,1} \nonumber\\
	&&-{\mathcal C}_{1212} 2 v^4_{,2}-{\mathcal C}_{1122} v^6_{,1}-{\mathcal C}_{1222} v^6_{,2} \nonumber\\
	&&-v^8_{,1}-v^{10}_{,2}=0
\end{eqnarray}

The sixth equation is compatibility equation which written with respect to $\rho_0 {\bf u}_t$ (after multiplying it with $\rho_0$) reads
\begin{eqnarray}
	&& \rho_0 {\mathcal C}_{1122} v^3_{,t} + 2 \rho_0 {\mathcal C}_{1222} v^4_{,t} + \rho_0 {\mathcal C}_{2222} v^6_{,t} \nonumber\\
	&&-{\mathcal C}_{1222} v^1_{,1} - {\mathcal C}_{2122} v^2_{,1} - {\mathcal C}_{1222} v^1_{,2} - {\mathcal C}_{2222} v^2_{,2}=0,
\end{eqnarray} 
which should be added to 
\begin{eqnarray}
	&&  {\mathcal C}_{1122} v^3_{,t} + 2  {\mathcal C}_{1222} v^4_{,t} +  {\mathcal C}_{2222} v^6_{,t} \nonumber\\
	&&-{\mathcal C}_{1222} v^5_{,1} - {\mathcal C}_{2122} v^7_{,1} - {\mathcal C}_{1222} v^5_{,2} - {\mathcal C}_{2222} v^7_{,2}=0,
\end{eqnarray} 
to give
\begin{eqnarray}
	&& (\rho_0+1) {\mathcal C}_{1122} v^3_{,t} + 2 (\rho_0+1) {\mathcal C}_{1222} v^4_{,t} + (\rho_0+1) {\mathcal C}_{2222} v^6_{,t} \nonumber\\
	&&-{\mathcal C}_{1222} v^1_{,1} - {\mathcal C}_{2122} v^2_{,1} - {\mathcal C}_{1222} v^1_{,2} - {\mathcal C}_{2222} v^2_{,2}\nonumber\\
	&&-{\mathcal C}_{1222} v^5_{,1} - {\mathcal C}_{2122} v^7_{,1} - {\mathcal C}_{1222} v^5_{,2} - {\mathcal C}_{2222} v^7_{,2}=0
\end{eqnarray} 

The seventh equation is momentum equation written with respect to ${\bf u}_t$ 
\begin{eqnarray}
	\rho_0 v^7_{,t}&&{-\mathcal C}_{2111} v^3_{,1}-{\mathcal C}_{2211} v^3_{,2}-{\mathcal C}_{2112} 2 v^4_{,1} \nonumber\\
	&&-{\mathcal C}_{2212} 2 v^4_{,2}-{\mathcal C}_{2122} v^6_{,1}-{\mathcal C}_{2222} v^6_{,2}=0,
\end{eqnarray}
which should be added to the momentum equation written with respect to the stresses
\begin{equation}
	\rho_0 v^7_{,t}-v^{10}_{,1}-v^9_{,2}=0,
\end{equation}
and these two last equations should be added to give
\begin{eqnarray}
	2 \rho_0 v^7_{,t}&&{-\mathcal C}_{2111} v^3_{,1}-{\mathcal C}_{2211} v^3_{,2}-{\mathcal C}_{2112} 2 v^4_{,1} \nonumber\\
	&&-{\mathcal C}_{2212} 2 v^4_{,2}-{\mathcal C}_{2122} v^6_{,1}-{\mathcal C}_{2222} v^6_{,2} \nonumber\\
	&& -v^{10}_{,1}-v^9_{,2}=0.
\end{eqnarray}

The three last equations are the constitutive laws time differentiated and written once with respect to $\rho_0 {\bf u}_t$ and once written with respect to ${\bf u}_t$. The first time differentiated constitutive law with respect to ${\bf u}_t$ reads
\begin{equation}
	{\mathcal C}^{-1}_{1111} v^8_{,t} + 2 {\mathcal C}^{-1}_{1112} v^{10}_{,t}+{\mathcal C}^{-1}_{1122} v^9_{,t}-v^5_{,1}=0,
\end{equation}
while the same equation with respect to $\rho_0 {\bf u}_t$ reads
\begin{equation}
	\rho_0 {\mathcal C}^{-1}_{1111} v^8_{,t} + 2 \rho_0 {\mathcal C}^{-1}_{1112} v^{10}_{,t}+\rho_0 {\mathcal C}^{-1}_{1122} v^9_{,t}-v^1_{,1}=0,
\end{equation}
so, by adding the last two equations we find
\begin{equation}
	{\mathcal C}^{-1}_{1111} (\rho_0 + 1) v^8_{,t} + 2 {\mathcal C}^{-1}_{1112} (\rho_0 + 1) v^{10}_{,t}+{\mathcal C}^{-1}_{1122} (\rho_0 + 1) v^9_{,t}-v^5_{,1}-v^1_{,1}=0.
\end{equation}

The ninth equation written with respect to ${\bf u}_t$ takes the form
\begin{equation}
	{\mathcal C}^{-1}_{2211} v^8_{,t} + 2 {\mathcal C}^{-1}_{2122} v^{10}_{,t}+{\mathcal C}^{-1}_{2222} v^9_{,t}-v^7_{,2}=0,
\end{equation}
while the same equation written with respect to $\rho_0 {\bf u}_t$ reads
\begin{equation}
	\rho_0 {\mathcal C}^{-1}_{2211} v^8_{,t} + 2 {\mathcal C}^{-1}_{12122} \rho_0 v^{10}_{,t} + \rho_0 {\mathcal C}^{-1}_{2222} v^9_{,t}-v^2_{,2}=0,
\end{equation}
so if we add the last two equations we find
\begin{equation}
	{\mathcal C}^{-1}_{2211} ( \rho_0 + 1 ) v^8_{,t} + 2 ( \rho_0 + 1 ) {\mathcal C}^{-1}_{2122} v^{10}_{,t} + ( \rho_0 + 1 ) {\mathcal C}^{-1}_{2222} v^9_{,t}-v^7_{,2}-v^2_{,2}=0.
\end{equation}

The tenth and last equation written with respect to ${\bf u}_t$ takes the form
\begin{equation}
	2 {\mathcal C}^{-1}_{1211} v^8_{,t} + 4 {\mathcal C}^{-1}_{2112} v^{10}_{,t} + 2 {\mathcal C}^{-1}_{1222} v^9_{,t}-v^5_{,2}-v^7_{,1}=0,
\end{equation}
while the same equation when written with respect to $\rho_0 {\bf u}_t$ takes the form 
\begin{equation}
	2 {\mathcal C}^{-1}_{1211} \rho_0 v^8_{,t} + 4 {\mathcal C}^{-1}_{2112} \rho_0 v^{10}_{,t} + 2 \rho_0 {\mathcal C}^{-1}_{1222} v^9_{,t}-v^1_{,2}-v^2_{,1}=0,
\end{equation}
and if we add these last two equations we take
\begin{equation}
	2 {\mathcal C}^{-1}_{1211} (\rho_0+1) v^8_{,t} + 4 {\mathcal C}^{-1}_{2112} (\rho_0+1) v^{10}_{,t} + 2 {\mathcal C}^{-1}_{1222} (\rho_0+1) v^9_{,t}-v^5_{,2}-v^7_{,1}-v^1_{,2}-v^2_{,1}=0.
\end{equation}

So, as a system of the form 
\begin{equation}
	A_{\text{Sym}}^{(\rho_0 {\bf u}_t, {\bf u}_t, {\bf e}, {\boldsymbol \sigma})} \frac{\partial {\bf q}}{\partial t} + \sum_{i=1}^2 B_{i \ \text{sym}}^{(\rho_0 {\bf u}_t, {\bf u}_t, {\bf e}, {\boldsymbol \sigma})} \frac{\partial {\bf q}}{\partial x_i}={\bf 0},
\end{equation} 
eqs. (155, 158, 161, 164, 167, 170, 173, 176, 179, 182) have matrices
{\setcounter{MaxMatrixCols}{20}
	\begin{equation}
		\rotatebox{90}{$ A_{\text{sym}}^{(\rho_0 {\bf u}_t, {\bf u}_t, {\bf e}, {\boldsymbol \sigma})}=\begin{pmatrix}
				2 & 0 & 0 & 0 & 0 & 0 & 0 & 0 & 0 & 0 \\
				0 & 2 & 0 & 0 & 0 & 0 & 0 & 0 & 0 & 0 \\
				0 & 0 & {\mathcal C}_{1111} (\rho_0+1) & 2{\mathcal C}_{2111} (\rho_0+1)  & 0 & {\mathcal C}_{2211}(\rho_0+1) & 0 & 0 & 0 & 0 \\
				0 & 0 & 2{\mathcal C}_{1112}(\rho_0+1) & 4{\mathcal C}_{2112} (\rho_0+1)  & 0 & 2{\mathcal C}_{2212} (\rho_0+1) & 0 & 0 & 0 & 0 \\
				0 & 0 & 0 & 0 & 2 \rho_0 & 0 & 0 & 0 & 0 & 0 \\
				0 & 0 & {\mathcal C}_{1122} (\rho_0+1) & 2 {\mathcal C}_{1222} (\rho_0+1) & 0 & {\mathcal C}_{2222} (\rho_0+1) & 0 & 0 & 0 & 0 \\
				0 & 0 & 0 & 0 & 0 & 0 & 2 \rho_0 & 0 & 0 & 0 \\ 
				0 & 0 & 0 & 0 & 0 & 0 & 0 & {\mathcal C}^{-1}_{1111} (\rho_0+1) & {\mathcal C}^{-1}_{1122} (\rho_0+1)  & 2{\mathcal C}^{-1}_{1112} (\rho_0+1) \\
				0 & 0 & 0 & 0 & 0 & 0 & 0 & {\mathcal C}^{-1}_{2211} (\rho_0+1) & {\mathcal C}^{-1}_{2222} (\rho_0+1) & 2{\mathcal C}^{-1}_{2212} (\rho_0+1) \\
				0 & 0 & 0 & 0 & 0 & 0 & 0 & 2{\mathcal C}^{-1}_{1211} (\rho_0+1) & 2{\mathcal C}^{-1}_{1222} (\rho_0+1)  & 4{\mathcal C}^{-1}_{1212} (\rho_0+1)
			\end{pmatrix},$}
\end{equation}}
which is symmetric when major and minor symmetries are enforced,
{\setcounter{MaxMatrixCols}{20}
	\begin{equation}
		\rotatebox{90}{$ 	B_{1 \ \text{sym}}^{(\rho_0 {\bf u}_t, {\bf u}_t, {\bf e}, {\boldsymbol \sigma})}=\begin{pmatrix}
				0 & 0  & -{\mathcal C}_{1111} &  -2{\mathcal C}_{1112} & 0 & -{\mathcal C}_{1122} & 0 & -1 & 0 & 0 \\
				0 & 0  & -{\mathcal C}_{2111} & -2{\mathcal C}_{2112} & 0 & -{\mathcal C}_{2122} & 0 & 0 & 0 & -1\\
				-{\mathcal C}_{1111} & -{\mathcal C}_{2111} & 0 & 0 & -{\mathcal C}_{1111} & 0 & -{\mathcal C}_{1211} & 0 & 0 & 0 \\
				-2 {\mathcal C}_{1112} & -2{\mathcal C}_{2112} & 0 & 0 & -2{\mathcal C}_{1112} & 0 & -2 {\mathcal C}_{2122} & 0 & 0 & 0\\
				0 & 0 & -{\mathcal C}_{1111} & -2{\mathcal C}_{1112} & 0 & -{\mathcal C}_{1222} & 0 & -1 & 0 & 0  \\
				-{\mathcal C}_{1222} & -{\mathcal C}_{2122} & 0 & 0 & -{\mathcal C}_{1222} & 0 & -{\mathcal C}_{2122} & 0 & 0 & 0 \\ 
				0 & 0 & -{\mathcal C}_{2111} & -2{\mathcal C}_{2122} & 0 & -{\mathcal C}_{2122} & 0 & 0 & 0 & -1 \\
				-1 & 0 & 0 & 0 & -1 & 0 & 0 & 0 & 0 & 0\\
				0 & 0 & 0 & 0 & 0 & 0 & 0 & 0 & 0 & 0 \\
				0 & -1 & 0 & 0 & 0 & 0 & -1 & 0 & 0 & 0 
			\end{pmatrix},$}
\end{equation}}
which is symmetric as well and
{\setcounter{MaxMatrixCols}{20}
	\begin{equation}
		\rotatebox{90}{$ 	B_{2 \ \text{sym}}^{(\rho_0 {\bf u}_t, {\bf u}_t, {\bf e}, {\boldsymbol \sigma})}=\begin{pmatrix}
				0 & 0  & -{\mathcal C}_{1211} & -2 {\mathcal C}_{1212} & 0 & -{\mathcal C}_{1222} & 0 & 0 & 0 & -1 \\
				0 & 0  & -{\mathcal C}_{2211} & -2 {\mathcal C}_{2212} & 0 & -{\mathcal C}_{2222} & 0 & 0 & -1 & 0 \\
				-{\mathcal C}_{1211} & -{\mathcal C}_{2211} & 0 & 0 & -{\mathcal C}_{1211} & 0 & -{\mathcal C}_{2211} & 0 & 0 & 0 \\
				-2 {\mathcal C}_{1212} & -2 {\mathcal C}_{2212} & 0 & 0 & -2 {\mathcal C}_{1212} & 0 & -2 {\mathcal C}_{2212} & 0 & 0 & 0 \\
				0 & 0 & -{\mathcal C}_{1211} & -2{\mathcal C}_{1212} & 0 & -{\mathcal C}_{1222} & 0 & 0 & 0 & -1 \\
				-{\mathcal C}_{1222} & -{\mathcal C}_{2222} & 0 & 0 & -{\mathcal C}_{1222} & 0 & -{\mathcal C}_{2222} & 0 & 0 & 0 \\
				0 & 0 & -{\mathcal C}_{2211} & -2{\mathcal C}_{2212} & 0 & -{\mathcal C}_{2222} & 0 & 0 & -1 & 0 \\ 
				0 & 0 & 0 & 0 & 0 & 0 & 0 & 0 & 0 & 0 \\
				0 & -1 & 0 & 0 & 0 & 0 & -1 & 0 & 0 & 0 \\
				-1 & 0 & 0 & 0 & -1 & 0 & 0 & 0 & 0 & 0 
			\end{pmatrix},$}
\end{equation}}
which is also symmetric.

\section{Conclusion}

The main achievement of the present contribution is that it offers alternative writing of the equations of classical elastodynamics in a symmetric format. For the one dimensional case we present symmetric writings with respect to variables $(u_t, u_x)$, $(u_t, \sigma)$, ($u_t, u_x, \sigma$), $(\rho_0 u_t, u_x)$, $(\rho_0 u_t, \sigma)$, ($\rho_0 u_t, u_x, \sigma$) and ($\rho_0 u_t, u_t, u_x, \sigma$). For the two dimensional case we present symmetric writings with respect to $({\bf u}_t, {\bf e})$, (${\bf u}_t, \boldsymbol \sigma$), (${\bf u}_t, {\bf e}, {\boldsymbol \sigma}$) and ($\rho_0 {\bf u}_t, {\bf u}_t, {\bf e}, {\boldsymbol \sigma}$). Granted that writing a system in symmetric format makes it amenable to computational approaches (\cite{Hughes-Tedbnyar}), the present writings may offer alternative paths for finite elements and this highlights the importance of the present contribution.

Such writings of the equations as a first order system enable us doing quite a lot of things for going forward: \\
i) seek the appropriate Hamiltonian structure and the associated conservation laws with each writing (\cite{Medvedev-Grebenev}), \\
ii) study the associated semigroup with each writing (\cite{Wilkes}), \\      
iii) in the nonlinear case there are alternative writings of the elasticity equations in a symmetric format (\cite{Boillat-Ruggeri,Qin1998}). It is perhaps intresting to search the lienarizations of such writings if and to which of the present writings corresponds. 

All in all, the present contribution's significance is that it presents alternative symmetric writings of the classical equations of elastodynamics, not given in the literature, to the best of our knowledge.

\section*{Acknowledgments}

The paper received no funding. The author declare that he has no conflict of interest.


%
%



\end{document}